\documentclass[floatfix,aps,pra,showpacs,twoside,twocolumn,10pt]{revtex4-2}
\usepackage[colorlinks=true, citecolor=red, urlcolor=blue ]{hyperref}
\usepackage{epsfig,newlfont,amssymb,amsfonts,amsmath,bm,subfigure,palatino,mathtools,amsthm,braket,times,soul,enumitem,color}
\usepackage[normalem]{ulem}
\usepackage{xcolor}

\usepackage{blindtext}
\usepackage{graphicx}
\usepackage{amsmath}
\usepackage{bm}
\usepackage{hyperref}
\usepackage{geometry}
\usepackage{amsthm}
\usepackage{physics}
 \definecolor{darkgreen}{rgb}{0.47,0.67,0.19}
\begin{document}



\title{Spread and asymmetry of typical quantum coherence\\and their inhibition in response to glassy disorder}

\author{George Biswas}
\author{Santanu Sarkar}
\author{Anindya Biswas}
\affiliation{Department of Physics, National Institute of Technology Sikkim, Ravangla, South Sikkim 737 139, India}
\author{Ujjwal Sen}
\affiliation{Harish-Chandra Research Institute,  A CI of Homi Bhabha National Institute, Chhatnag Road, Jhunsi, Prayagraj 211 019, India}

\begin{abstract}
We consider  the average quantum coherences of typical redits and qudits - vectors of real and complex Hilbert spaces - with the analytical forms stemming from the symmetry of Haar-uniformly distributed random pure states. We subsequently study the response to disorder in spread of the typical quantum coherence in response to glassy disorder. The disorder is inserted in the state parameters. Even in the absence of disorder, the quantum coherence distributions of redits and qudits are not uniform over the range of quantum coherence, and the spreads are lower for higher dimensions. On insertion of disorder, the spreads decrease.  This decrease in the spread of quantum coherence distribution in response to disorder is seen to be a generic feature of typical pure states: we observe the feature for different strengths of disorder and for various types of disorder distributions, viz.  Gaussian, uniform, and Cauchy-Lorentz. We also find that the quantum coherence distributions become less asymmetric with increase in dimension and with infusion of glassy disorder.
\end{abstract}

\maketitle

\section{Introduction}
\label{qc_int}
The phenomenon of quantum coherence emerges due to the ``waveness'' of quantum systems. 
Quantum entanglement~\cite{RevModPhys.81.865,GUHNE20091,das2017separability,doi:https://doi.org/10.1002/9783527805785.ch8}
, another quintessential quantum phenomenon, appears in  shared states of at least two quantum systems, whereas quantum coherence appears due to interference of at least two quantum waves associated with a single system.
While the concept of quantum coherence was known since the beginnings of quantum theory, a formal apparatus of the resource theory of quantum coherence is rather recent~\cite{aberg2006quantifying,PhysRevLett.113.140401}
. 
There has been a 
large number of studies on different aspects of the resource theory, including Refs.~\cite{PhysRevA.92.022124,PhysRevLett.115.020403,PhysRevLett.116.120404,PhysRevLett.116.150502,PhysRevA.93.012110,Qi_2017,PhysRevA.96.042336,PhysRevA.96.052336,RevModPhys.89.041003,PhysRevLett.119.230401,PhysRevLett.123.110402,PhysRevResearch.1.033020,dey2019structure,Das_2020,PhysRevA.103.022417,PhysRevA.103.032429,banerjee2021quantum}
. Along with its fundamental importance in the structure of quantum theory of physical systems, quantum coherence has also been connected to a variety of other physical phenomena, see e.g.~\cite{Scully2011-od,PhysRevA.86.043843,Horodecki2013,Skrzypczyk2014,Abah_2014,PhysRevLett.112.030602,Faist_2015,vacanti2015work,PhysRevE.92.042150,Lostaglio2015,PhysRevA.93.052335,Korzekwa_2016,Kammerlander2016,PhysRevLett.116.150502,PhysRevA.93.042107,PhysRevA.93.012111,anand2016coherence,Matera_2016,Matera_2016}
.

The importance of considering the response of disorder in a physical phenomenon can hardly be over-emphasized. Disorder appears naturally and almost universally in physical systems.
{The system under study, when realized, is expected to be affected by several such processes. Their individual effects can be of varied forms, and they may arise due to an array of physical effects, e.g., stray fields or imperfect system elements like a randomly-appearing defect in a lattice~\cite{PhysRevA.96.033606}. In our study, we have modeled these effects by incorporating a disordered state parameter (or a few of them) in the system state.
The disorder itself} may appear in different hues and arrangements in a substrate realizing a phenomenon. We will be interested in a category of disorder that has been called ``glassy'' or ``quenched'' in the literature~\cite{doi:10.1142/0223,doi:10.1142/0271,Chakrabarti1996,nishimori01,sachdev_2011,Suzuki2013}
, and refers to a disorder in which a system parameter picks up values randomly from a certain distribution, decided by its preparation procedure and physical nature, and the typical  equilibration time of the disorder is far greater than the typical system manipulation times that we are interested in. 
{We therefore assume that the physical effects that lead to the disordered parameter are ``slow'' with respect to the timescales over which we observe our system. After we have chosen the relative equilibration timescale of the disorder, we still need to decide on the distribution of the disorder, and this will depend on the actual physical realization of the physical system. In the hope that the results obtained are somewhat generic, we consider three probability distributions for the disorder, one of which does not have a well-defined mean.}

In this paper, we focus on the distribution of quantum coherence of Haar-random pure states, and the response of the same when the state parameters are inflicted by glassy disorder. {Random pure states are the quantum analog of random numbers in classical information theory, and this is the reason that we chose to look at the effect of disorder in random pure states, as it would then provide us with a ``typical'' response to the disorder.}
In the following section (Sec.~\ref{int_sub}), we present an introductory discussion on the Haar uniformity of random pure states and the distribution functions used during the introduction of disorder in state parameters. 
In 
Sec.~\ref{qc_sec_2}, we present the analytical forms of the average quantum coherences (as quantified by the $l_1$-norm of quantum coherence~\cite{aberg2006quantifying,PhysRevLett.113.140401}
) of Haar uniform random pure states, for both ``redits'' and ``qudits''. In Sec.~\ref{qc_sec_3}, we discuss the variation of  quantum coherence distributions obtained for Haar-uniformly generated pure states for  Hilbert spaces of different dimensions. In Sec.~\ref{qc_sec_4}, we analyze the effect of disorder in the state parameters on the distributions of Haar-uniformly generated pure states. A conclusion is presented in Sec.~\ref{qc_sec_5}.

\section{Haar uniformity and probability distribution functions}\label{int_sub}
In this paper, we study the quantum coherence of typical real bits and quantum bits, and their \(d\)-dimensional versions, for Hilbert spaces of  different dimensions. 
The quantum coherences are always considered in the 
computational basis. 
 A \(d\)-dimensional pure quantum state 
 can be written as 
\begin{equation} \label{bit_int}
    \ket{\psi}=\sum_{j=1}^{d}(c_{1j}+ic_{2j})\ket{j},
\end{equation}
where $\ket{j}$ represents the \(j^{\text{th}}\) orthonormal basis vector in the computational basis of  the $d$-dimensional complex Hilbert space, \(\mathbb{C}^d\), and $c_{1j},~c_{2j}$ are real  numbers, 
constrained by the normalization condition, \(\langle \psi | \psi \rangle = 1\). It is called a quantum bit or qubit for $d=2$, qutrit for $d=3$, and qudit in general.
If the imaginary parts of coefficients of the basis vectors in Eq.~(\ref{bit_int}) are vanishing (\textit{i.e.}, if $c_{2j}=0$ \(\forall j\)), then the corresponding vector can be called a real bit or rebit for $d=2$, a  retrit for $d=3$, and a redit in general~\cite{Caves2001,Wootters2012}. A real bit has, in general,  off-diagonal terms in the computation basis, and so is different from a probabilistic mixture of being in two orthogonal states, although both can be parametrized by a single real number. ``Quantum coherence'' is typically defined for states of or defined on a complex Hilbert space. We however will use the same definition and \emph{terminology}, also for states of or defined on real Hilbert spaces. 

Haar uniformity is attained by choosing the \(c_{ij}\) independently from 
Gaussian distributions with vanishing mean and 
finite  variance~\cite{Press92numericalrecipes,bengtsson_zyczkowski_2006,cohn2013measure,Dahlsten_2014,Biswas_2021}. 
The obtained state will have to be normalized to unity. 
The probability density function for a Gaussian distribution is given by 
\begin{equation} \label{gaussian_int}
f_G(x)=\frac{1}{\sigma_G\sqrt{2\pi}}e^{-\frac{1}{2}\left(\frac{x-\mu_G}{\sigma_G}\right)^2},
\end{equation}
where $\mu_G$ is the mean and $\sigma_G$ is the standard deviation of the distribution. Before normalization, the states generated are distributed in a hyperspace of dimension $d$ when the real numbers are selected from the Gaussian distribution with zero mean and finite variance.
After normalization, they are distributed uniformly over a hypersphere of unit radius in the same hyperspace.
Note that the joint probability distribution of 
independent variables is spherically symmetric if the individual distributions are Gaussian.

{Random generation of quantum states involves choosing the coefficients of the states from independent Gaussian distributions. We have used the Gaussian distribution to generate Haar uniform random pure states. We also used the Gaussian distribution} in the process of insertion of disorder in those states. Beside Gaussian disorder we introduce disorder from uniform and Cauchy-Lorentz distributions as well. The uniform distribution may be represented by the probability density function
\begin{equation} \label{e3}
    f_U(x)= 
\begin{cases}
    \frac{1}{b-a},& \text{if } a\leq x\leq b,\\
    0,              & \text{otherwise.}
\end{cases}
\end{equation}
The mean, standard deviation, and semi-interquartile range for the uniform distribution in terms of \(a\) and \(b\), that marks the terminal points of the non-trivial part of the uniform distribution, are given respectively by 
\begin{equation}
\mu_U=\frac{b+a}{2} \quad \text{,} \quad \sigma_U=\frac{b-a}{2\sqrt{3}}, \quad \text{and} \quad \gamma_U=\frac{b-a}{4}.
\end{equation}


The probability density function of the uniform distribution can be rewritten in terms of its mean and standard deviation as
\begin{equation} \label{e5}
    f_U(x)= 
\begin{cases}
    \frac{1}{2\sqrt{3}\sigma_U},& \text{if } \mu_U-\sqrt{3}\sigma_U\leq x\leq \mu_U+\sqrt{3}\sigma_U,\\
    0,              & \text{otherwise.}
\end{cases}
\end{equation}
\par The Cauchy-Lorentz probability density function 
is given by
\begin{equation} \label{e7}
f_{C-L}(x|x_0,\gamma_{C-L})=\frac{\gamma_{C-L}}{\pi\left[\gamma_{C-L}^2+\left(x-x_0\right)^2\right]},
\end{equation}
where $x_0$ is the median of the distribution and $\gamma_{C-L}$ is the 
semi-interquartile range of the distribution. The Cauchy-Lorentz distribution has ``long tails'', as a result of which the mean and variance do not exist. The Cauchy principal value of the mean is well-defined, and
is equal to the median of the distribution.  
One may use the median 
as a measure of the central tendency of this distribution. In the absence of the standard deviation, the semi-interquartile range may be used as a measure of dispersion of the distribution. A useful function is the cumulative distribution function of the Cauchy-Lorentz distribution, given by 
\begin{align} \label{e8}
F_{C-L}(x|x_0,\gamma_{C-L})&=\int_{-\infty}^{x}f_{C-L}(x'|x_0,\gamma_{C-L})\dd{x'} \nonumber \\
&=\frac{1}{\pi}{\tan}^{-1}\left(\frac{x-x_0}{\gamma_{C-L}}\right)+\frac{1}{2}.
\end{align}

The quantile function or the inverse cumulative distribution function is given by
\begin{equation} \label{e9}
x=x_0+\gamma_{C-L}\tan\left[\pi\left(F_{C-L}-\frac{1}{2}\right)\right].
\end{equation}
A random number from the Cauchy-Lorentz distribution is obtained using this quantile function when $F_{C-L}$ is randomly chosen from a uniform distribution in the range 0 to 1.




\section{Average quantum coherence of typical pure states}
\label{qc_sec_2}

A Haar-random qubit in computational basis can be written as 
\begin{equation}
\label{quantum_bit}
|q_2\rangle=(c_{11}+ic_{21})\begin{pmatrix} 1\\0 \end{pmatrix}+(c_{12}+ic_{22})\begin{pmatrix} 0\\1 \end{pmatrix}.
\end{equation}
The random coefficients $c_{11}$, $c_{21}$, $c_{12}$, $c_{22}$ are chosen independently from a Gaussian distribution of mean zero and finite variance, and the state must then be normalized. {We use a computable measure of quantum coherence, namely $l_1$-norm, given by}

\begin{equation} \label{L1n_unmdfd}
C_{l_1}(\rho)=\sum_{i\neq j}| {\rho}_{ij}|,
\end{equation}
{where \({\rho}_{ij}\) is the \((i,j)^{\text{th}}\) matrix element of the state \(\rho\) in the computational basis.}
 
 {We analytically calculate the average $l_1$-norm of typical qubits exploiting the spherical symmetry of Haar uniform random states.} The algebra is similar to that of the four-dimensional redit, considered below, and leads to 
\begin{equation} \label{l1n_qu2}
    \overline{C}_{l_1}(|q_2\rangle) = \frac{\pi}{4}.
\end{equation}

We have performed the calculations for higher-dimensional qudits also. {Note that} it has been analytically shown in Ref.~\cite{bu2016average}, by following an elegant mathematical procedure, that 
\(\overline{C}_{l_1}(|q_d\rangle) = (d-1)\frac{\pi}{4}\).
{Also note that the similar general method for arbitrary dimensional redits, as shown in Eq.~(\ref{l1n_red}), is also applicable to calculate the average $l_1$-norm of arbitrary dimensional qudits.}

We now move over to real bits. A redit is represented by
\begin{equation}\label{n_redit}
|r_d\rangle=\sum_{j=1}^d c_j |j\rangle,
\end{equation}
where $c_i$s are real numbers and $|j\rangle$  represents the $j^{\text{th}}$ orthonormal basis vector in the computational basis of
the \(d\)-dimensional real Hilbert space. The corresponding density matrix is given by $$\rho_{r_d}=\begin{bmatrix} c_1^2 & ... & ... & c_1c_n \\ ...&...&...&... \\ c_nc_1& ...&... & c_n^2 \end{bmatrix}.$$

Let us first deal with a real bit or a rebit. A Haar-random rebit in the computational basis may be written as
\begin{equation}
\label{real_bit}
|r_2\rangle=c_1\begin{pmatrix} 1\\0 \end{pmatrix}+c_2\begin{pmatrix} 0\\1 \end{pmatrix},
\end{equation}
where $c_1$ and $c_2$ are random real numbers. The corresponding density matrix is $\rho_{r_2}=\begin{bmatrix} c_1^2 & c_1c_2 \\ c_2c_1 & c_2^2 \end{bmatrix}$. The random coefficients $c_1$ and $c_2$ are chosen independently from a Gaussian distribution of mean zero, and then the state is normalized.
The Haar-uniformly chosen rebits are uniformly distributed over the circumference of the unit circle.
The $l_1$-norm of quantum coherence of the rebit is given by $C_{l_1}(|r_2\rangle)=2|c_1c_2|$. In the polar coordinate system, with \(c_1=\cos\theta\) and \(c_2=\sin\theta\), the $l_1$-norm becomes 
\begin{equation}
\begin{split}
    C_{l_1}(|r_2\rangle) &= 2|\cos\theta \sin\theta|\\
    &= |\sin2\theta|
\end{split}
\end{equation}

The average of the $l_1$-norms of quantum coherence of these Haar uniformly distributed rebits is given by 

\begin{equation} \label{l1n_re2}
\begin{split}
    \overline{C}_{l_1}(|r_2\rangle) &= \frac{\int_0^{\pi/2} \sin2\theta d\theta}{\int_0^{\pi/2} d\theta}\\
    &= \frac{2}{\pi}.\\
\end{split}
\end{equation}

We choose the integration limits of \(\theta\) as $0\leq \theta \leq \pi/2$ to avoid keeping track of the signs in the sines and cosines in a larger interval. As the distribution of the rebits is uniform over the circumference of the unit circle, the averaging over the $1^{\text{st}}$ quadrant and the same  over the whole circle are equivalent. {Following a similar procedure, we find \(C_{l_1}(|r_3\rangle)=\frac{4}{\pi} \text{~and~} C_{l_1}(|r_4\rangle)=\frac{6}{\pi}\).} 

{The average $l_1$-norm of quantum coherence for higher-dimensional redits can be calculated in closed form using the same procedure as discussed above.}
{For arbitrary dimension `$d$', 
$C_{l_1}(|r_d\rangle=2\abs{\sum_{j=1}^{d}c_j c_{j+1}}$ with $c_{d+1}=c_1$. We need to convert the coefficients of basis vectors to the unit polar coordinate system. We require $d-1$ angles, and say, they are $\{\theta_1, \theta_2,...,\theta_{d-1}\}$. The transformation will be \begin{equation} \label{polar_d}
\begin{split}
    c_1 &=\cos\theta_1,\\
    c_2 &=\sin\theta_1 \cos\theta_2,\\
    c_3 &=\sin\theta_1 \sin\theta_2 \cos\theta_3,\\
    :\\
   c_{d-1} &=\sin\theta_1 \sin\theta_2 ... \cos\theta_{d-1},\\
    c_d &=\sin\theta_1 \sin\theta_2 ... \sin\theta_{d-1}.\\
\end{split}
\end{equation} 
The surface element of the $d$-dimensional unit sphere is $ds_d=\sin^{d-2}\theta_1 \sin^{d-3}\theta_2 ... \sin\theta_{d-2} d\theta_1 d\theta_2 ... d\theta_{d-1}$. Similar to the previous cases, the average $l_1$-norm of quantum coherence of Haar uniformly distributed $d$-dimensional redits is given by 
\begin{equation} \label{l1n_red}
\begin{split}
    \overline {C}_{l_1}(|r_d\rangle) &= \frac{\displaystyle\int_{\theta_1=0}^{\pi/2} \int_{\theta_2=0}^{\pi/2} ... \int_{\theta_{d-1}=0}^{\pi/2} \{C_{l_1}(|r_d\rangle)\} ds_d}{\displaystyle\int_{\theta_1=0}^{\pi/2} \int_{\theta_2=0}^{\pi/2} ... \int_{\theta_{d-1}=0}^{\pi/2} ds_d}.\\
\end{split}
\end{equation} 
The above integration has been done numerically upto dimension $d=6$. The result for average $l_1$-norm} can be checked in even higher dimensions using numerically simulated Haar-uniform states, and this we have checked until $d=10$, and we propose that for a redit, \(\overline{C}_{l_1}(|r_d\rangle)    = (d-1)\frac{2}{\pi}\). The Haar-uniform simulation is discussed in Sec.~\ref{qc_sec_2}, and the numerical integration is performed using Gaussian quadratures.

{Real bits or rebits are the basic ingredients of the two-dimensional real vector space quantum mechanics~\cite{https://doi.org/10.48550/arxiv.quant-ph/9905037}. It has been shown that quantum computation can be performed using rebits and gates that belong to real vector space~\cite{Adleman97quantumcomputability,https://doi.org/10.48550/arxiv.quant-ph/0210187,PhysRevX.5.021003}. The entanglement properties of rebits have also been investigated~\cite{Caves2001,Batle2003,Wootters_2014}. Recently, the entanglement of two rebits  has been realized in laboratories~\cite{PhysRevA.103.L040402,Sperling:21}. Redits are qudits with real coefficients. Real vector space is a subset of complex vector space, so redits are a subset of qudits. In this paper, we study the quantum coherence properties of typical redits and typical qudits. We find that their behavior in response to disorder is qualitatively similar that of qubits.}

\section{Distribution of quantum coherence of typical pure states}
\label{qc_sec_3}
In the preceding section, we have calculated the average $l_1$-norms of quantum coherence for typical redits and qudits. The maximum values of these $l_1$-norms (for redits and qudits) depend on the dimension $d$ of the vector space, and is given by $d-1$.
In this section, we present a comparative study of average quantum coherences of typical pure states for different dimensions. For the comparative analysis, we modify the definition of the $l_1$-norm so that for every dimension, it lies between 
zero and unity.
The modified 
$l_1$-norm of quantum coherence of a density matrix $\rho_d$, acting on a \(d\)-dimensional Hilbert space, is  therefore
\begin{equation} \label{L1n_modified}
C_{l_1}^m(\rho_d)=\frac{1}{d-1}\sum_{i\neq j}| {\rho_d}_{ij}|=\frac{C_{l_1}(\rho_d)}{d-1},
\end{equation}
where \({\rho_d}_{ij}\) is the \((i,j)^{\text{th}}\) matrix element of \(\rho_d\) in the computational basis. 

\begin{figure}[!ht]
  \centering
    \includegraphics[width=\linewidth]{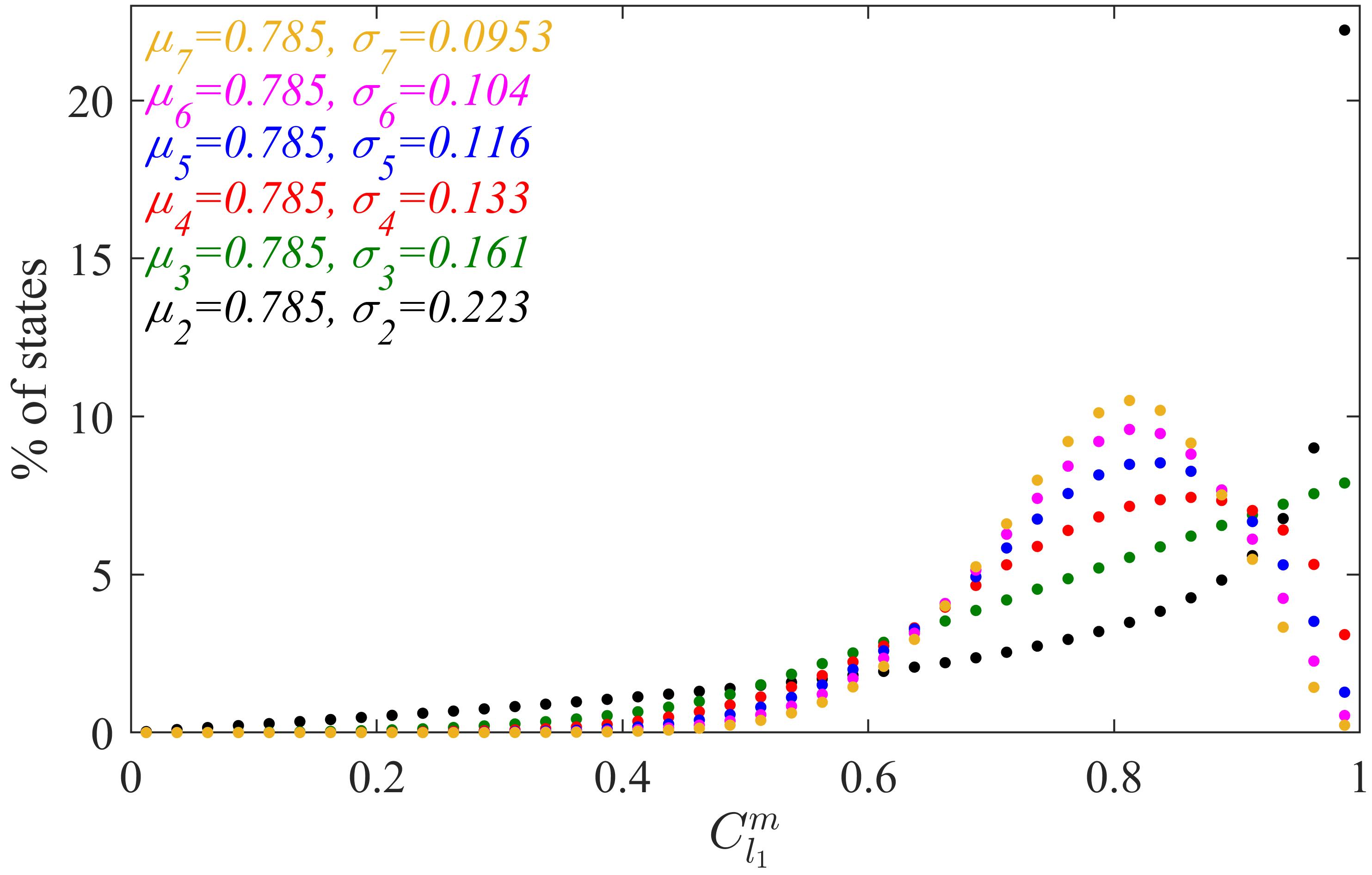}
    \caption{Distribution of typical quantum coherence. We plot the relative frequency percentages of Haar-uniformly chosen qudits against the modified $l_1$-norm values, which lie between $0$ and $1$. The black dots correspond to relative frequency percentages of random qubit pure states generated Haar uniformly. The green dots correspond to relative frequency percentages of Haar-uniformly generated random qutrit pure states. The red, blue, magenta, and orange dots correspond to qudit pure states for $d=4,5,6,7$ respectively. The mean ($\mu_d$) and standard deviation ($\sigma_d$) for different $d$ are tabulated on the figure. The calculation of the relative frequency percentages requires a window on the horizontal axis, and it is chosen as 0.025. The Haar-uniform generation used for the figure utilizes a million states. The plots do not alter up to the precision used for 10 million states, and the precision is checked to three significant figures. The skewnesses of the plots are given by $s_2=-1.15$, $s_3=-0.865$, $s_4=-0.737$, $s_5=-0.661$, $s_6=-0.598$, and $s_7=-0.561$. All quantities are dimensionless.
    }
  \label{l1n_qubit_ordered}
\end{figure}

\begin{figure}[!ht]
  \centering
    \includegraphics[width=\linewidth]{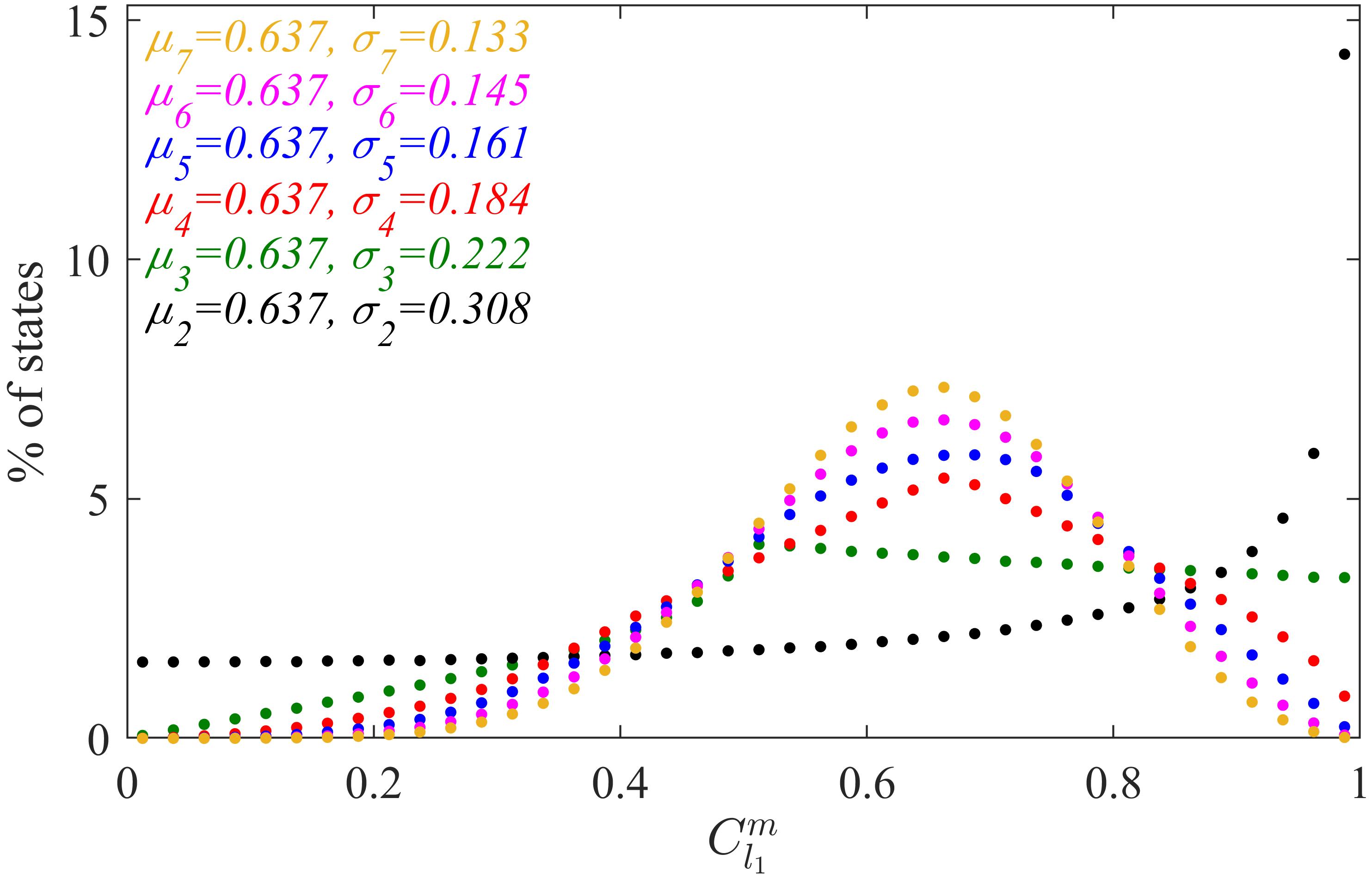}
    \caption{Distribution of quantum coherence of typical redits. We plot the relative frequency percentages of Haar-uniformly chosen redits against the modified $l_1$-norm values, which lie between $0$ and $1$. The considerations are the same as in Fig.~\ref{l1n_qubit_ordered}, except the fact that we used redits instead of qudits. 
    The skewnesses of the plots are given by $s_2=-0.498$, $s_3=-0.375$, $s_4=-0.327$, $s_5=-0.302$, $s_6=-0.282$, and $s_7=-0.269$. 
    All quantities used are dimensionless.
    }
  \label{l1n_rebit_ordered}
\end{figure}

{After this normalization of the $l_1$-norm of quantum coherence the average of the modified $l_1$-norm $(C_{l_1}^m(\rho_d))$ of typical qudits is $\frac{\pi}{4}\approx 0.785$ and of typical redits is $\frac{2}{\pi}\approx 0.637$ for any dimension. The numerically calculated mean values of typical qudits and typical redits as written in the insets of  Fig.~\ref{l1n_qubit_ordered} and Fig.~\ref{l1n_rebit_ordered} are in agreement.} 


 An arbitrary qudit pure state is given by Eq.~(\ref{bit_int}).
It 
can be  Haar-uniformly generated by randomly choosing the real parameters in the coefficients  independently from  a Gaussian distribution of mean $\mu_G = 0$ and a finite standard deviation, 
 followed by a  normalization~\cite{Zyczkowski_2001,doi:10.1063/1.3595693,MISZCZAK2012118,PhysRevA.93.032125,e20100745,Dahlsten_2014,cohn2013measure,Biswas_2021}. Here we draw the random numbers independently from the Gaussian distribution with mean $\mu_G = 0$ and standard deviation $\sigma_G = 1$.
 
 A million such random pure states are generated, and the modified $l_1$-norm (Eq.~(\ref{L1n_modified})) of each state is calculated. Thus, a quantum coherence distribution for random qudit pure states is obtained, ranging from 0 to 1. The relative frequency percentages of the distribution are plotted in Fig.~\ref{l1n_qubit_ordered}.
 
 The average value of the modified $l_1$-norm, as obtained in our numerics, remains fixed at \(\approx 0.785\) for every dimension, which is nearly equal to the analytical result $\frac{\pi}{4}$, given in Sec.~\ref{qc_sec_2}. The distribution however hides further information. We quantify the spread of typical quantum coherences for different dimensions by the corresponding standard deviations and skewnesses.

 For a data set with data points, $\{(x_1,y_1), (x_2,y_2), \ldots, (x_N,y_N)\}$, the skewness \(s\) is defined as 
\begin{equation} \label{skewness_definition}
s=\dfrac{\sum_{i=1}^{N}{(y_i-\mu)^3}}{N \sigma^3},
\end{equation}
where $\mu$ and $\sigma$ are the mean and the standard deviation of the 
data set. Skewness measures the asymmetry of the distribution. It is negative when the distribution has a longer left tail and positive for a distribution with a longer right tail~\cite{Hippel2011}. 

\begin{figure}[!ht]
\centering

\textbf{\textcolor{black}{(a)}}
\includegraphics[width=7.7cm]{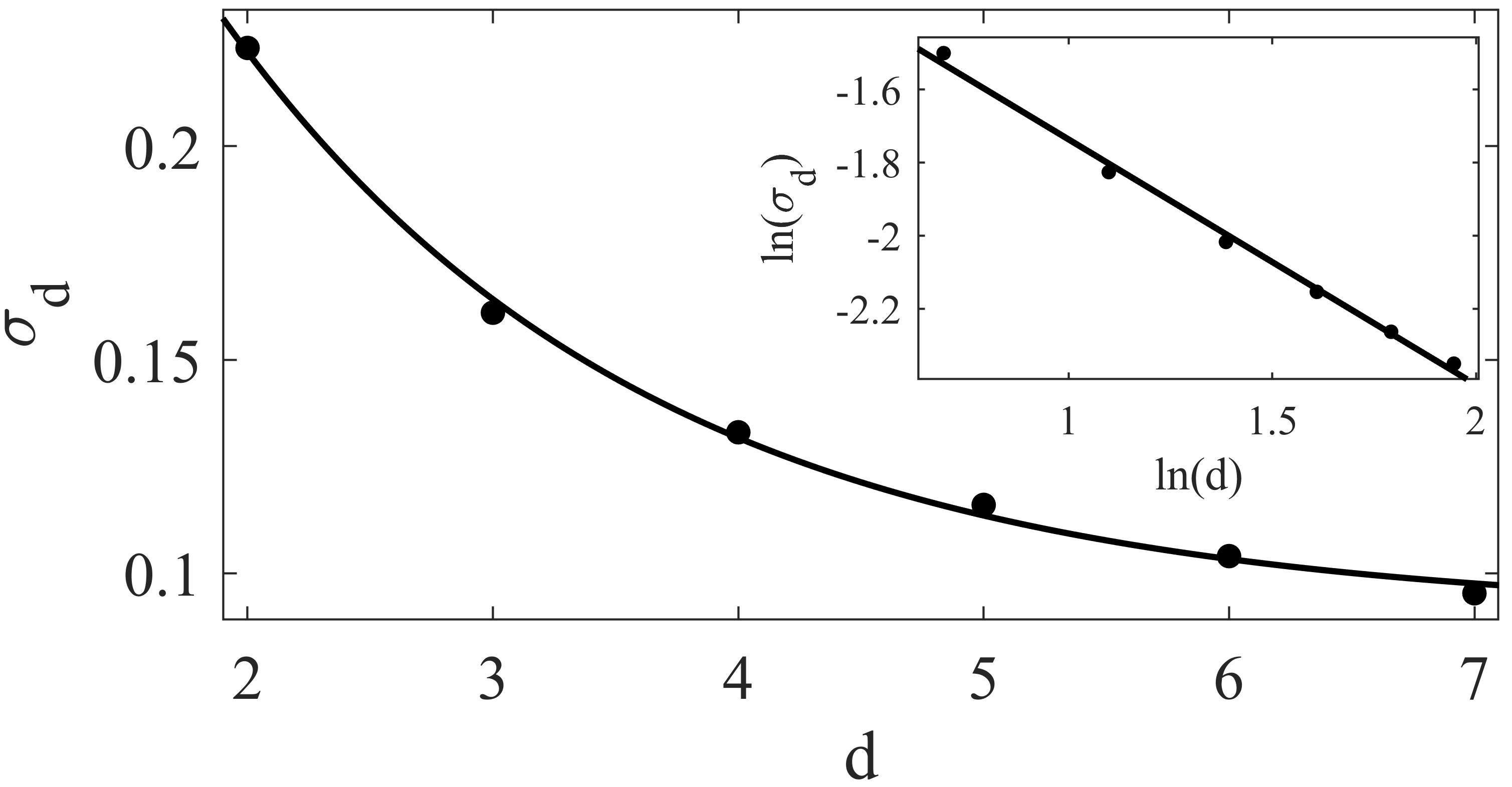}

\textbf{\textcolor{blue}{(b)}}
\includegraphics[width=7.7cm]{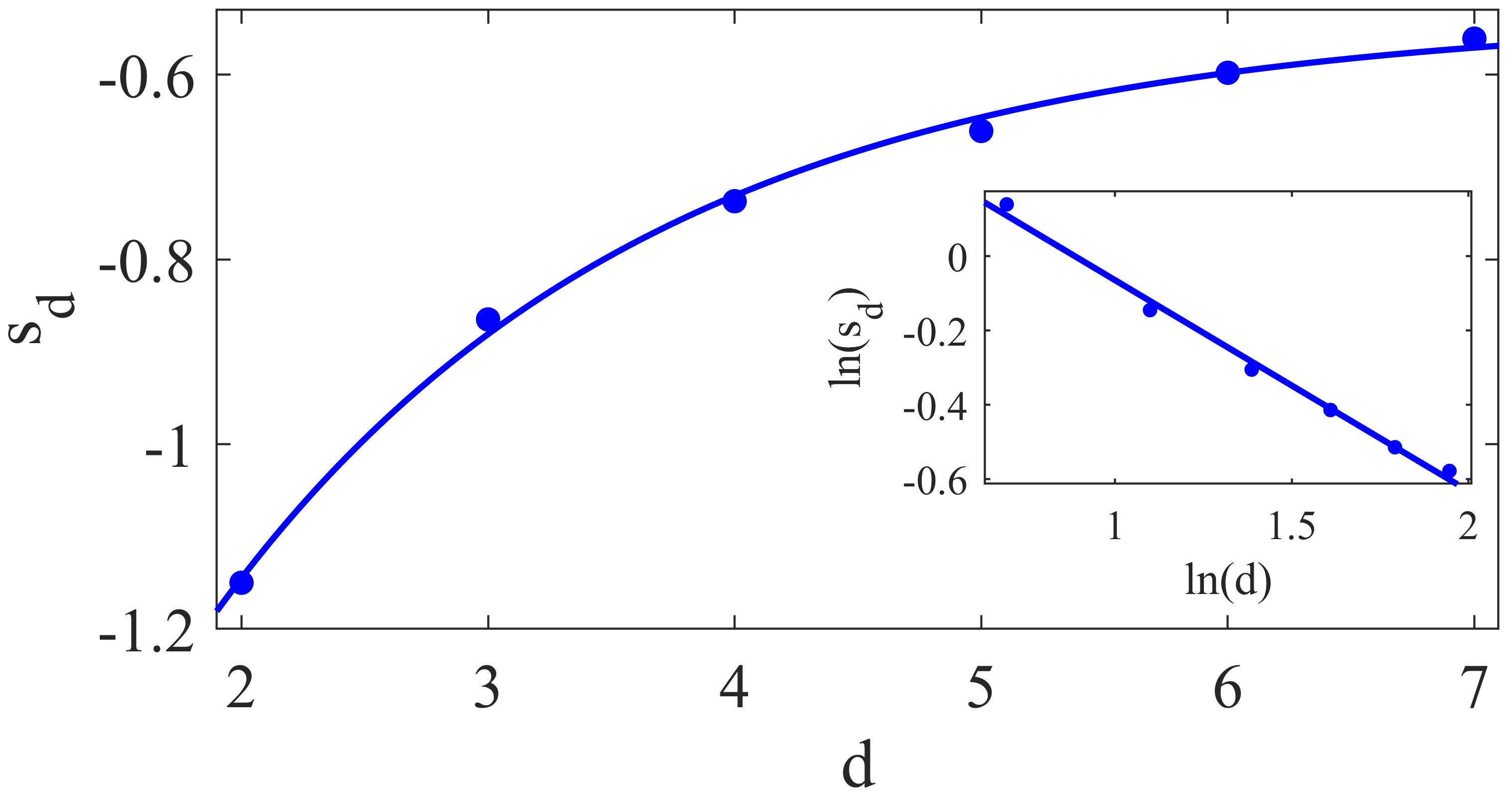}

\caption{Variation with dimension of spread and asymmetry of typical quantum coherence distributions. We plot the spread, as quantified by standard deviation (\(\sigma_d\)), and asymmetry, as quantified by skewness (\(s_d\)), of the different typical quantum coherence distributions for different dimensions, against the dimension (\(d\)). Both cases can be fitted well with exponential curves, \(f(d:\alpha,\beta,\gamma) = \alpha e^{-\beta d} + \gamma\). In (a), the exponential curve is given by \(\sigma_d = f(d:\alpha,\beta,\gamma)\), with \(\alpha = 0.419\), \(\beta = 0.579\), \(\gamma = 0.0903\).  
The goodness of fit, as measured by the sum of squares of errors, is given by $4.08\times 10^{-6}$. In \textcolor{blue}{(b)}, the exponential  is given  
by \(s_d = f(d:\alpha,\beta,\gamma)\), with \(\alpha = -1.90\), \(\beta = 0.567\), \(\gamma = -0.535\). 
The goodness of fit in this case is $1.04\times 10^{-4}$. {The inset plots are in the log-log scale.} All quantities used are dimensionless.
}
\label{damp_exp_qubit}
\end{figure}
 
We find that on increasing the dimension of the Hilbert space, the standard deviations and left skewnesses of the relative frequency percentage plots of Fig.~\ref{l1n_qubit_ordered}, i.e., of the spread of typical quantum coherence as measured by the modified \(l_1\)-norm of quantum coherence, decrease exponentially, as shown in  Fig.~\ref{damp_exp_qubit}. {These observations are in agreement with those reported in Ref.~\cite{bu2016average}, wherein it has been shown that 
the standard deviation of the distribution of the typical modified or normalized $l_1$-norm concentrates around its mean value and goes to zero for infinite dimension.}

\begin{figure}[!ht]
\centering

\textbf{\textcolor{black}{(a)}}
\includegraphics[width=7.7cm]{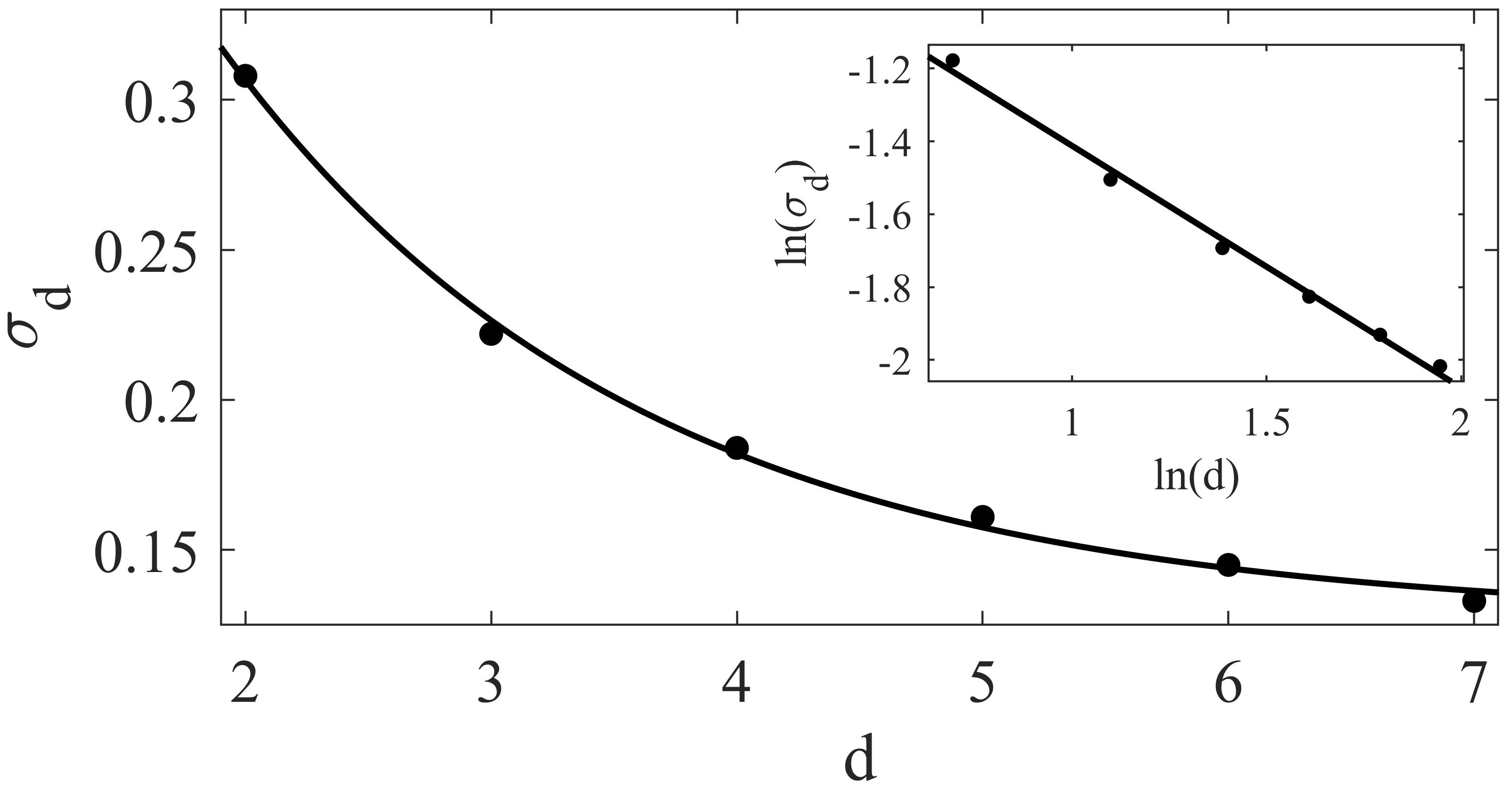}

\textbf{\textcolor{blue}{(b)}}
\includegraphics[width=7.7cm]{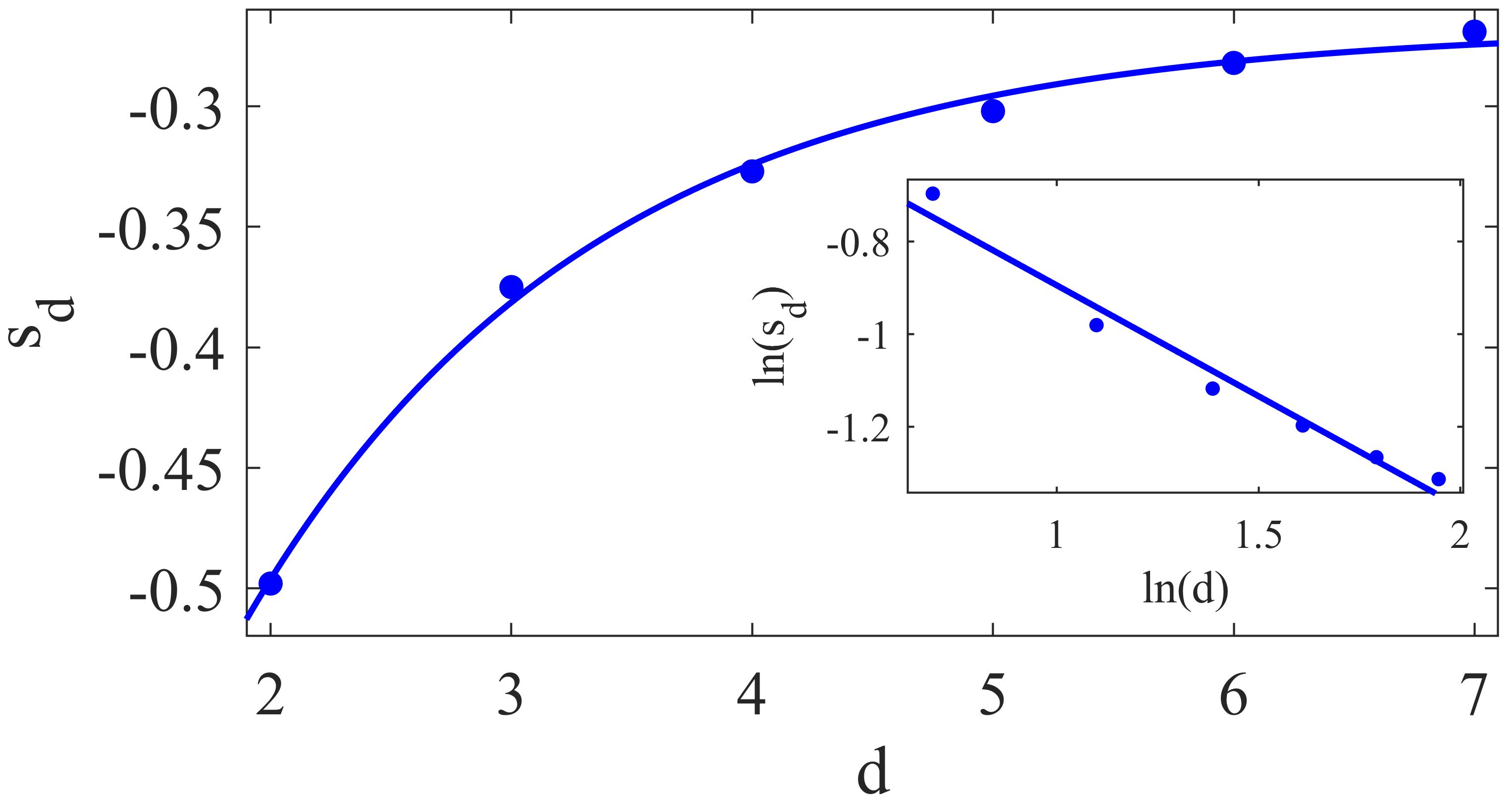}

\caption{Variation with dimension of spread and asymmetry of  quantum coherence distributions of typical redits. The considerations are the same as in Fig.~\ref{damp_exp_qubit}. The curves are again fitted with exponentials, although the parameters are different, and so in (a), here, the fitting function is 
\(\sigma_d = f(d:\alpha,\beta,\gamma)\), with \(\alpha = 0.586\), \(\beta = 0.591\), \(\gamma = 0.127\). 
The corresponding goodness of fit is $8.38\times 10^{-6}$. And in \textcolor{blue}{(b)}, the fitting function is \(s_d = f(d:\alpha,\beta,\gamma)\), with \(\alpha = -0.923\), \(\beta = 0.697\), \(\gamma = -0.267\). 
The goodness of fit is $2.02\times 10^{-5}$. {The inset plots are in the log-log scale.} All quantities used are dimensionless. 
}
\label{damp_exp_rebit}
\end{figure}

 Relative frequency percentages of redits for different dimensions are plotted in Fig.~\ref{l1n_rebit_ordered}. These curves are similar to the corresponding curves for qudits. The typical average value of the modified $l_1$-norm remains fixed at \(\approx 0.637\) for every dimension, which is nearly equal to the value  $\frac{2}{\pi}$, found in Section~\ref{qc_sec_2}. Similar to the case of qudits, here, in the case of redits also, the standard deviation and left skewness decrease exponentially with increasing  dimension of the Hilbert space, as depicted in Fig.~\ref{damp_exp_rebit}. The equations of the exponential curves, for both qudits and redits, are given in the corresponding figure captions.
 
 We therefore find that the distribution of typical quantum coherence of qudits as well as redits become more and more concentrated around their average values as we go towards higher dimensional systems. The distributions also become more and more symmetric with increasing dimension, from being left-skewed in low dimensional systems.

\begin{figure}[!ht]
  \centering
   \textbf{\textcolor{black}{(a)}}
\includegraphics[width=7.7cm]{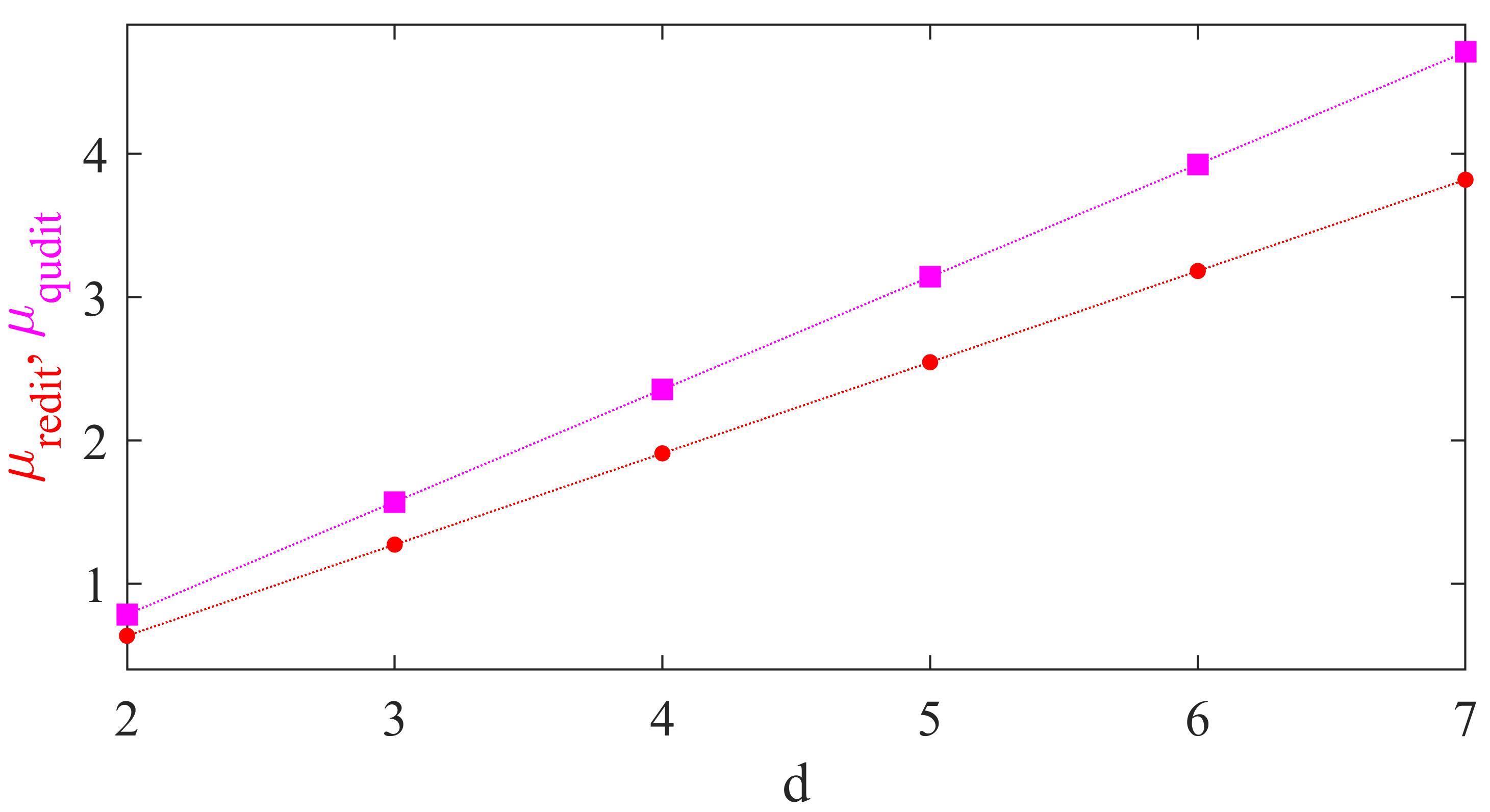}

\textbf{\textcolor{blue}{(b)}}
\includegraphics[width=7.7cm]{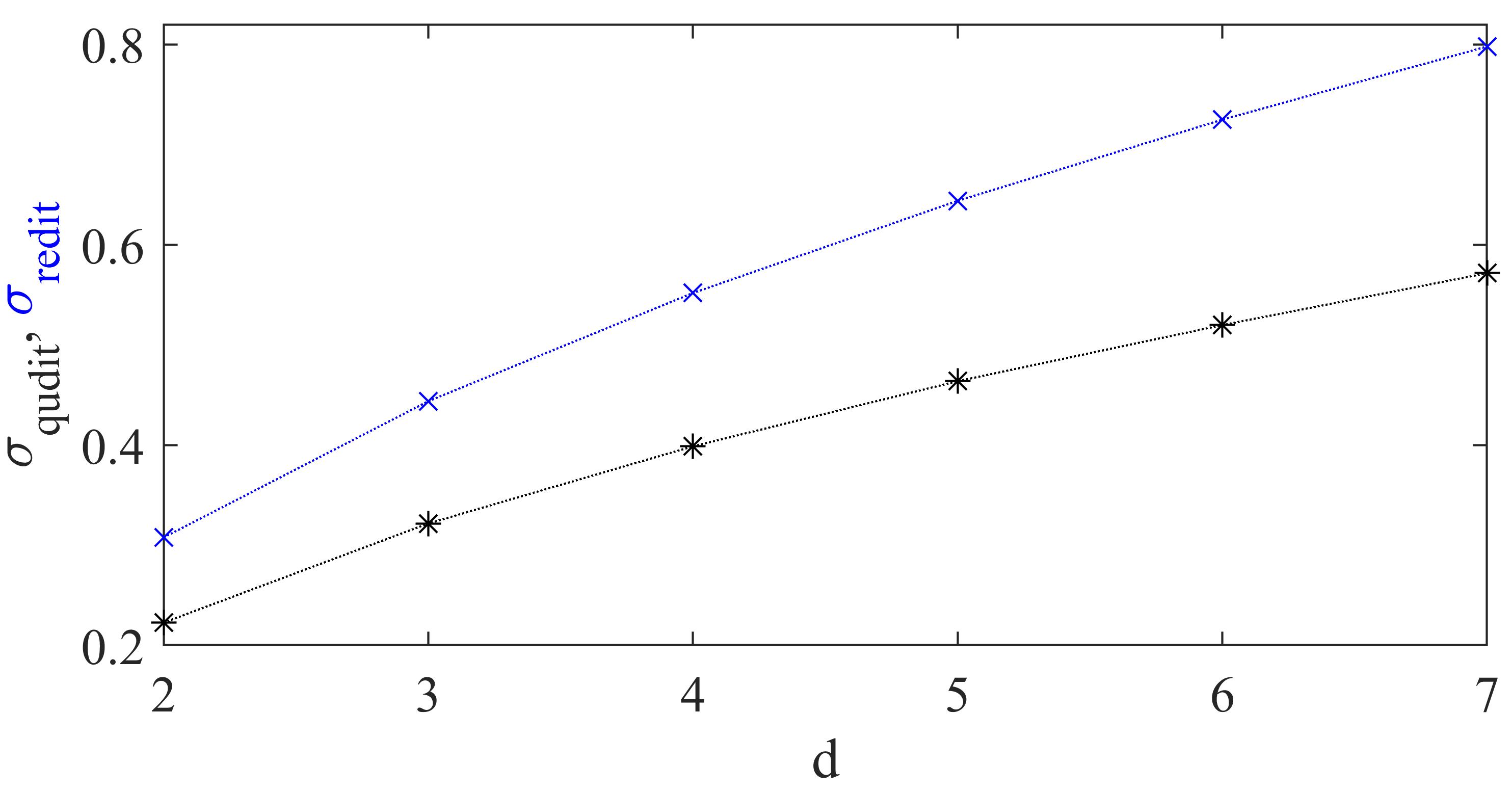}
    \caption{{Means (a) and standard deviations \textcolor{blue}{(b)} of the relative frequency percentages plots using un-normalized $l_1$-norm are plotted against the dimension for qudits and redits.}}
  \label{un_norm_mu_sigma}
\end{figure}

{The standard deviation of the distribution of the typical modified or normalized $l_1$-norm concentrate around its mean value and goes to zero as the dimension goes to infinity~Ref.~\cite{bu2016average}. Our observations agree with the results reported in~Ref.~\cite{bu2016average}. The concentration of the typical modified or normalized $l_1$-norm around its mean value for higher dimensions is not an artifact of normalization. Note that the un-normalized average $l_1$-norm of typical Haar uniform random qudits increase linearly with the increase in dimension of the vector space, real or complex, while the un-normalized standard deviations of the relative frequency percentage distributions  increase gradually at a slower rate with the dimension of the vector space (see Fig.~\ref{un_norm_mu_sigma}). 
The normalized or the modified $l_1$-norm is essential to compare the quantum coherence distributions of different dimensional qudits and redits.}

\section{Effect of disorder on the spread of quantum coherence of typical pure states}
\label{qc_sec_4}
We now investigate the response of the distribution of quantum coherence for typical states on introduction of disorder in the parameters of the states. 
We 
introduce disorder in  real parts of the coefficients $c_{1j}$, using 
Gaussian, uniform, or  Cauchy-Lorentz distributions.
We begin with the case of Gaussian disorder.

\begin{figure}[!ht]
  \centering
    \includegraphics[width=\linewidth]{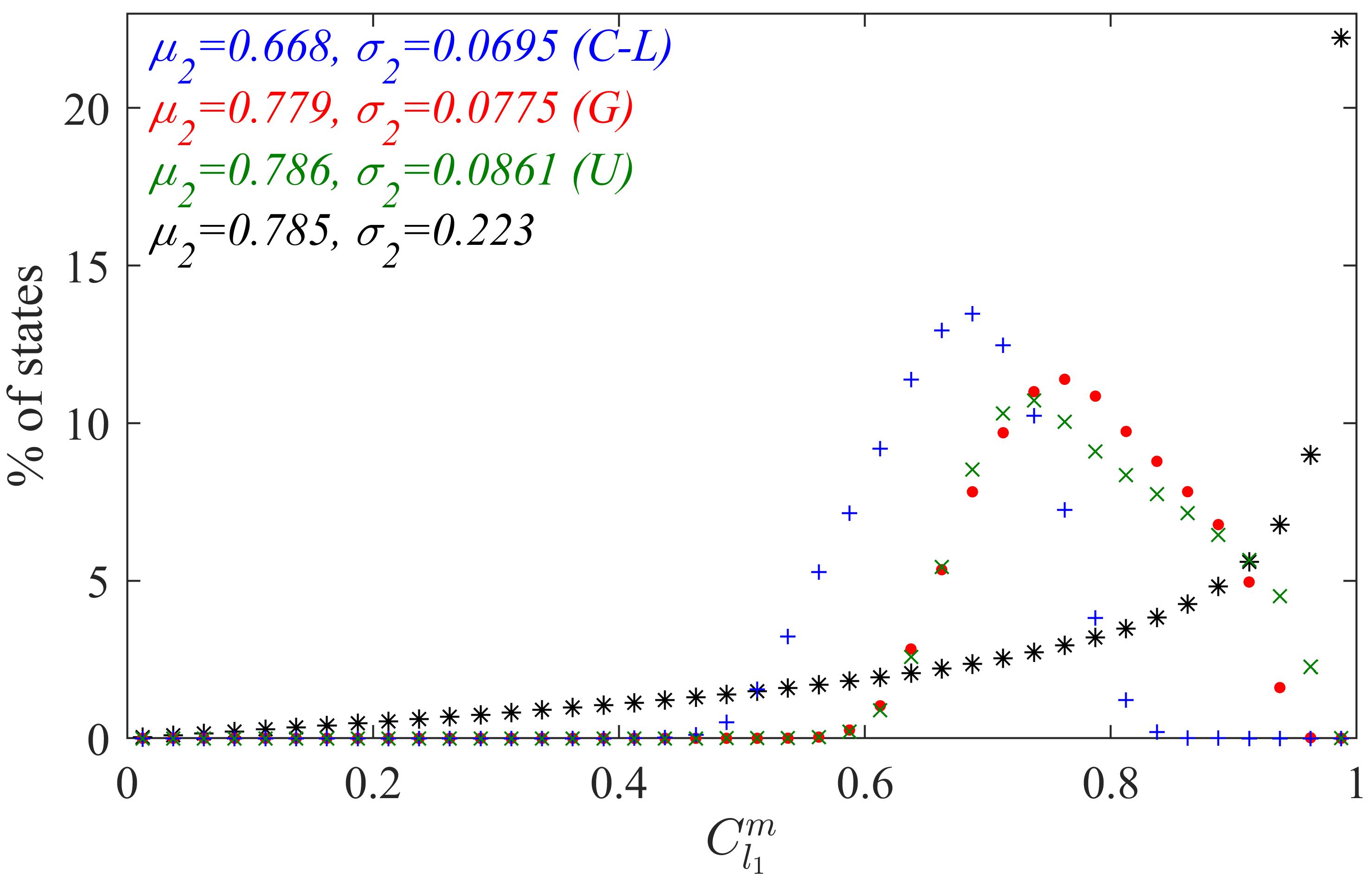}
    \caption{Inhibition of spread of quantum coherence of typical qubits in response to disorder. We plot here the relative frequency percentages of Haar-uniformly chosen
    pure qubits, with and without disorder. The disorder, whenever present, is in the real parts of all the Fourier coefficients of the states in the computational basis. This relative frequency is plotted  against the modified $l_1$-norm values, with the latter lying between $0$ and $1$. The black asterisks correspond to percentages of random pure qubits generated Haar uniformly, and without any disorder inflicted. The other three curves depict quantum coherence distributions for random pure qubits with the disorder  chosen from Gaussian (\(G\), red dots), uniform (\(U\), green crosses), and Cauchy-Lorentz (\(C-L\), blue pluses) distributions. The disorder-averaged values plotted in the figure are for \(100\) disorder configurations for every ordered one. The disorder-averaged curves does not change up to the precision used for averaging over \(50\) configurations. The initial Haar-uniform generation used for the figure utilizes \(10^6\) states. However, the same plot does not alter, up to the precision used, for \(10^7\) states. The precision is checked to three significant figures. The skewnesses for the ordered case and the disordered cases from uniform, Gaussian, and Cauchy-Lorentz distributions are
    $s=-1.15$, 
    $s_U=-0.177$, $s_G=0.0161$, and $s_{C-L}=-0.225$ respectively. The means (\(\mu_{2}\)) and standard deviations (\(\sigma_2\)) for the curves are given in the legend. All quantities used are dimensionless.}
  \label{l1n2}
\end{figure}

In the \emph{first step}, we generate a million Haar-random pure qudits. 
In the \emph{second step}, one hundred random pure states are generated for every state generated in the first step. It is in the second step that disorder is inserted. Each state \(\sum_{j=1}^{d}(\tilde{c}_{1j}+ic_{2j})\ket{j}\) in the second step is created by randomly choosing numbers, \(\tilde{c}_{1j}\), which are Gaussian distributed with mean  
$\mu_G = c_{1j}$, where $c_{1j}$ is the random number generated in the first step, and semi-interquartile-range $\gamma_G = 1/2$.
It may be noted that \(\gamma_G = 1/2\) corresponds to a standard deviation of  $\approx 0.741$. The  Gaussian distribution needed here to insert disorder should not be confused with the Gaussian distribution required in the context of Haar-uniform generation of states.
The standard deviation and semi-interquartile range of the Gaussian distribution used here quantifies the strength of the error or disorder inserted.
%
The \(c_{2j}\) in the second step are the same as those 
selected in the first step  of Haar uniform pure state generation after the normalization in the first step. The random pure states are normalized and their average quantum coherence calculated.
We end up with another set of one million 
disorder averaged quantum coherence values. We plot the corresponding relative frequency percentages for the  $C_{l_1}^m$ for qubits, as red dots in Fig.~\ref{l1n2}. For a comparative study between different distributions of disorder, we fix the semi-interquartile ranges of Gaussian, uniform, and Cauchy-Lorentz distributions at $\gamma_G=\gamma_U=\gamma_{C-L}=\frac{1}{2}$ while inserting disorders. The type of disorder we consider here is often referred to as ``quenched disorder'' in the literature (see e.g.~\cite{doi:10.1142/0223,doi:10.1142/0271,Chakrabarti1996,nishimori01,sachdev_2011,Suzuki2013}). The averaging needs to be performed after all physical quantities for given realizations of the disorder have been calculated. Such an averaging has often been referred to
as ``quenched averaging'' in the literature (see e.g.~\cite{Saha_1994,PhysRevE.72.061905,Blavatska_2013}).

 We see from Fig.~\ref{l1n2} that the relative frequency percentage plot  changes significantly with the introduction of disorder. The mean $C_{l_1}^m$ does not change much except when the disorder distribution is  Cauchy-Lorentz. The spread of the plot is reduced significantly - the standard deviation decreases to about a third of its ordered case value -  for every kind of disorder considered. The left skewnesses of the plots are also reduced, so that the disorder-averaged plots are more symmetric than the ordered one. The reduction in standard deviation is minimum for disorder from uniform distribution and maximum for the same from Cauchy-Lorentz distribution. 
The reduction in left skewness is minimum in the case of disorder from Cauchy-Lorentz distribution and maximum in the case of the same from Gaussian distribution. 

\begin{figure}[!ht]
  \centering
    \includegraphics[width=\linewidth]{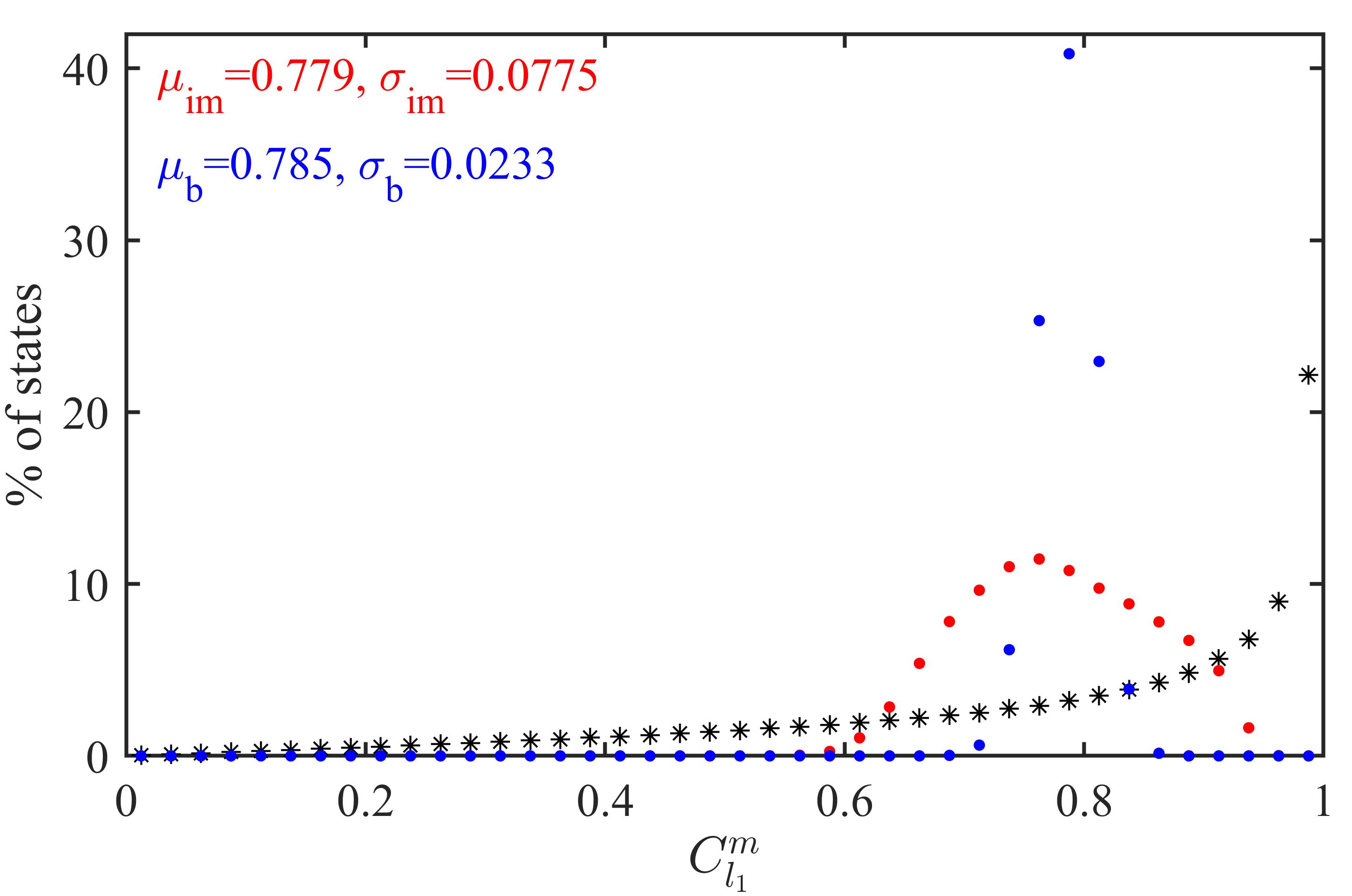}
    \caption{{Inhibition of spread of quantum coherence of typical qubits when the disorders are inserted in the imaginary parts of the Fourier coefficients. The considerations are the same as in Fig~\ref{l1n2}, except we consider Gaussian disorder in the imaginary parts only of all the Fourier coefficients in the red curve and Gaussian disorder in both real and imaginary parts of all the Fourier coefficients in the blue curve. The subscripts `im' and `b' are used in the inset to denote the disorder in imaginary parts and both parts, respectively.}}
  \label{ib}
\end{figure}

{Fig.~\ref{ib} shows the effect of Gaussian disorder on the relative frequency percentages of quantum coherence of qubits when we introduce the disorder in the imaginary parts of the Fourier coefficients as well. The red dots in Fig.~\ref{ib} represent the relative frequency percentages of states when we introduce Gaussian disorder \((\gamma_G=\frac{1}{2})\) in the imaginary parts of the Fourier coefficients only. The relative frequency percentage distribution is identical, as intuitively expected, to the case when we introduce Gaussian disorder \((\gamma_G=\frac{1}{2})\) in the real parts of the Fourier coefficients only, vide the red dots in Fig.~\ref{l1n2}. The relative frequency percentage distribution represented by the blue dots in Fig.~\ref{ib}, where we insert Gaussian disorder \((\gamma_G=\frac{1}{2})\) in both the real and imaginary parts of the Fourier coefficients, is also qualitatively similar to the plots where we introduce Gaussian disorder in the real or the imaginary parts only \textit{i.e.}, the standard deviation of the distribution is drastically reduced. In the rest of the paper, we analyze the effect of disorder by inserting them in the real parts of the Fourier coefficients only.}

\begin{figure}[!ht]
  \centering
    \includegraphics[width=\linewidth]{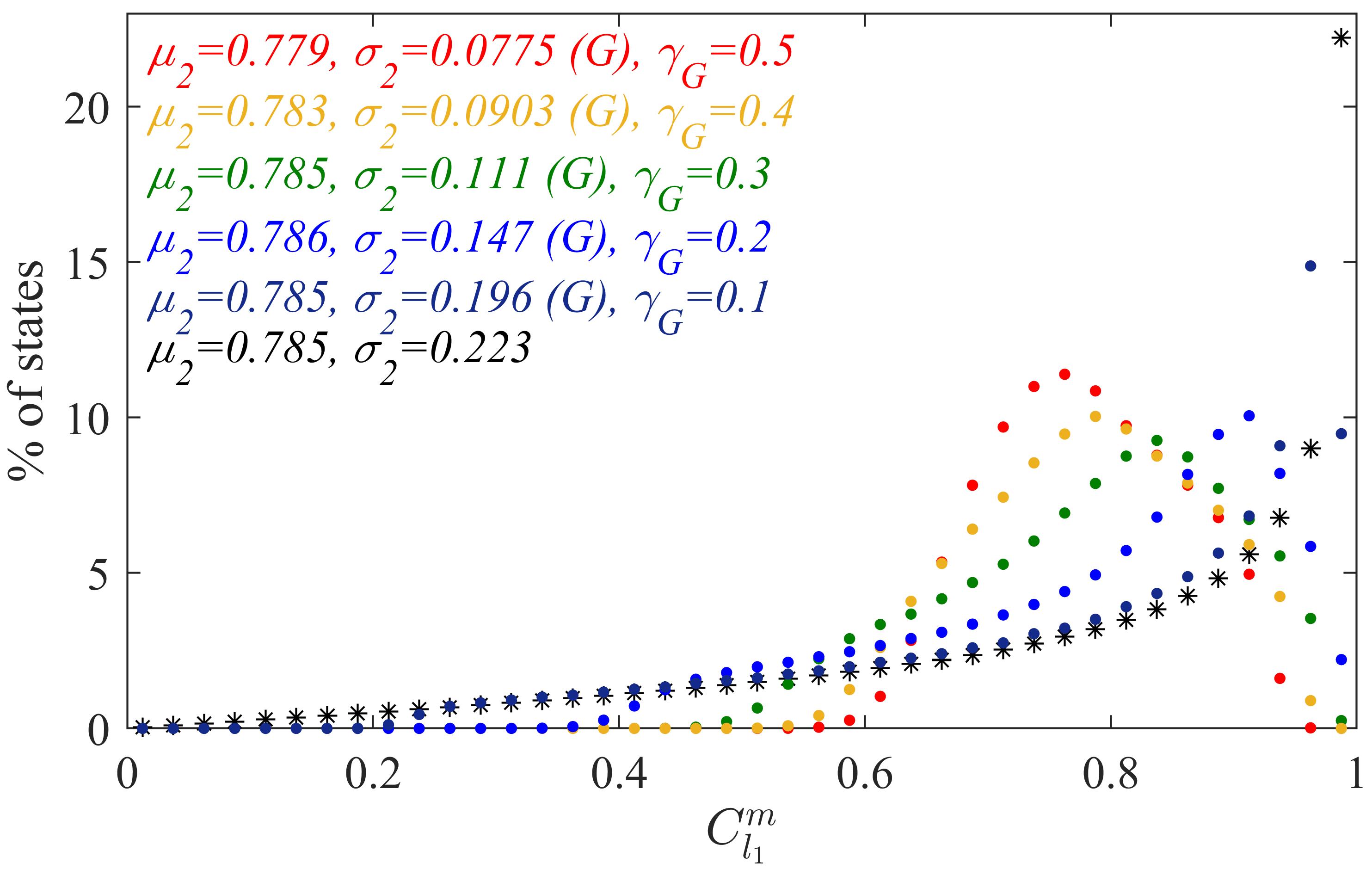}
    \caption{Effect of variation of the strength of Gaussian disorder. The considerations are the same as in Fig.~\ref{l1n2}, except that only the Gaussian disorder is considered. Different semi-interquartile ranges are used with different symbols for the plots. Black asterisks are used for the case when there is no disorder, and is also present in 
    Fig.~\ref{l1n2}. The curve with red dots is for \(\gamma_G = 1/2\) and again was also present in Fig.~\ref{l1n2}. The deep-blue, blue, green, and orange dots are for \(\gamma_G = 0.1\), \(0.2\), \(0.3\), and \(0.4\) respectively. Their skewnesses are $s_{0.1}=-1.00$, $s_{0.2}=-0.787$, $s_{0.3}=-0.462$, and $s_{0.4}=-0.158$ respectively. All quantities are dimensionless.
    }
  \label{qubit_siqr}
\end{figure}

Fig.~\ref{qubit_siqr} shows the effect of variation of the strength of Gaussian disorder on the relative frequency percentages of quantum coherence of qubits. We find that the spread of the distributions reduce with the increase of the strength of the disorder. 

\begin{figure}[!ht]
  \centering
    \includegraphics[width=\linewidth]{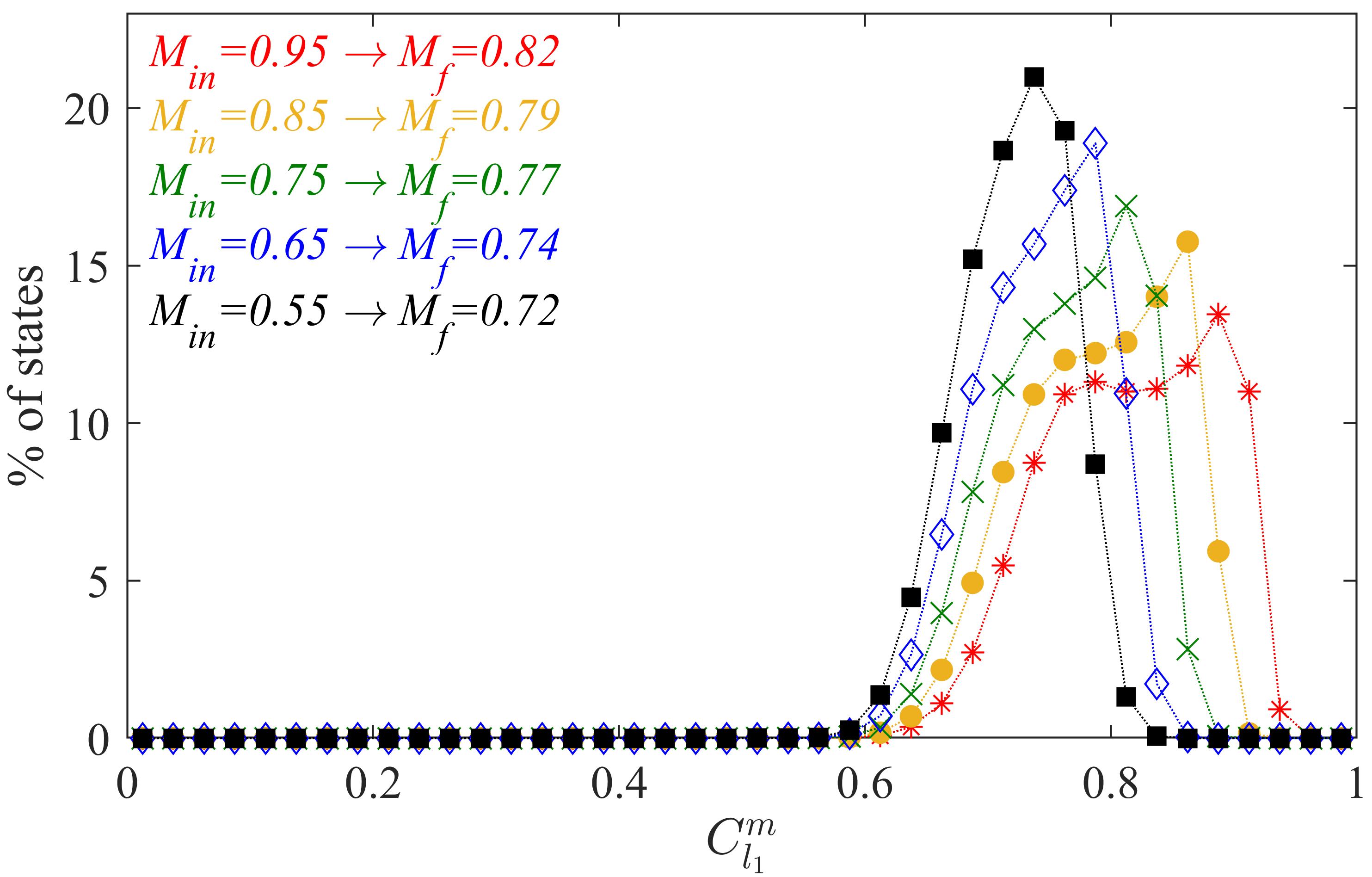}
    \caption{Response to disorder inserted in states of fixed quantum coherence. We analyze here the dependence of the disorder-averaged quantum coherence $M_f$ in response to Gaussian disorder, in the Fourier coefficients,
    on the initial quantum coherence $M_{in}$. We generate \(x=10^7\) Haar-uniform pure qubit states and choose all that have a 
    quantum coherence (in computational basis) within the window \((M_{in} - .01, M_{in}+.01)\). 
    For every such state, we generate \(y=100\) disordered 
    states by randomly choosing the real parts of the Fourier coefficients in the computational basis from a Gaussian distribution with mean equal to the input Fourier coefficients and \(\gamma_G=1/2\), and calculate the quantum coherence in the computational basis of the states. 
    We then plot the relative frequency  percentages of the states against the quantum coherences, with a window of 0.025.
    The red asterisks, orange dots, green crosses, blue rhombuses, and black squares respectively represent the cases for \(M_{in} = 0.95, 0.85, 0.75, 0.65, 0.55\). 
    The values written in the upper left corner are correct up to two significant figures. All quantities are dimensionless.
    To check convergence of the percentages obtained, we re-do the calculations with \(x=10^8\), and do not find any appreciable change. Similarly, we re-do the calculations for \(y=50\), and do not find a significant difference of the numbers within the precision considered.
    %
    }
  \label{qubit_ios}
\end{figure}

We have therefore found 
that the spread of quantum coherence of typical quantum states
is inhibited, seemingly irrespective of the type and strength of the disorder, and that the inhibition amount depends on the type and strength of disorder. To understand the reason for the inhibition of spread, we investigate further and introduce disorder in the same Fourier coefficients as before, but in qubit states with  specific values of quantum coherence, $M_{in}$, and look at the disorder-averaged  quantum coherence, \(M_f\). For the analysis, we will need a window of  quantum coherence for the input states, and we arbitrarily choose it to be \((M_{in}-.01, M_{in}+.01)\).
It can be seen~from Fig.~\ref{qubit_ios} that the ``final'' (i.e., disorder-averaged) quantum coherence $M_f$ is greater than the ``initial'' value of quantum coherence $M_{in}$, when $M_{in}< \frac{\pi}{4}$, whereas the order is inverted 
when $M_{in}> \frac{\pi}{4}$. Note that the average quantum coherence of typical qubit states, without any disorder, is \(\pi/4\). 
Therefore, it appears that as we perturb a quantum state which has a certain value of
$C_{l_1}^m$, it has a greater probability to transform into a state with higher or lower $C_{l_1}^m$, depending on whether the parent state had a value of $C_{l_1}^m$ that was lower or higher than the average $C_{l_1}^m$ of Haar uniformly distributed states in the ordered case. 
Note that 
the shift of average quantum coherence, $|M_f-M_{in}|$, increases with $|M_{in}-\frac{\pi}{4}|$. 

\begin{figure}[!ht]
  \centering
    \includegraphics[width=\linewidth]{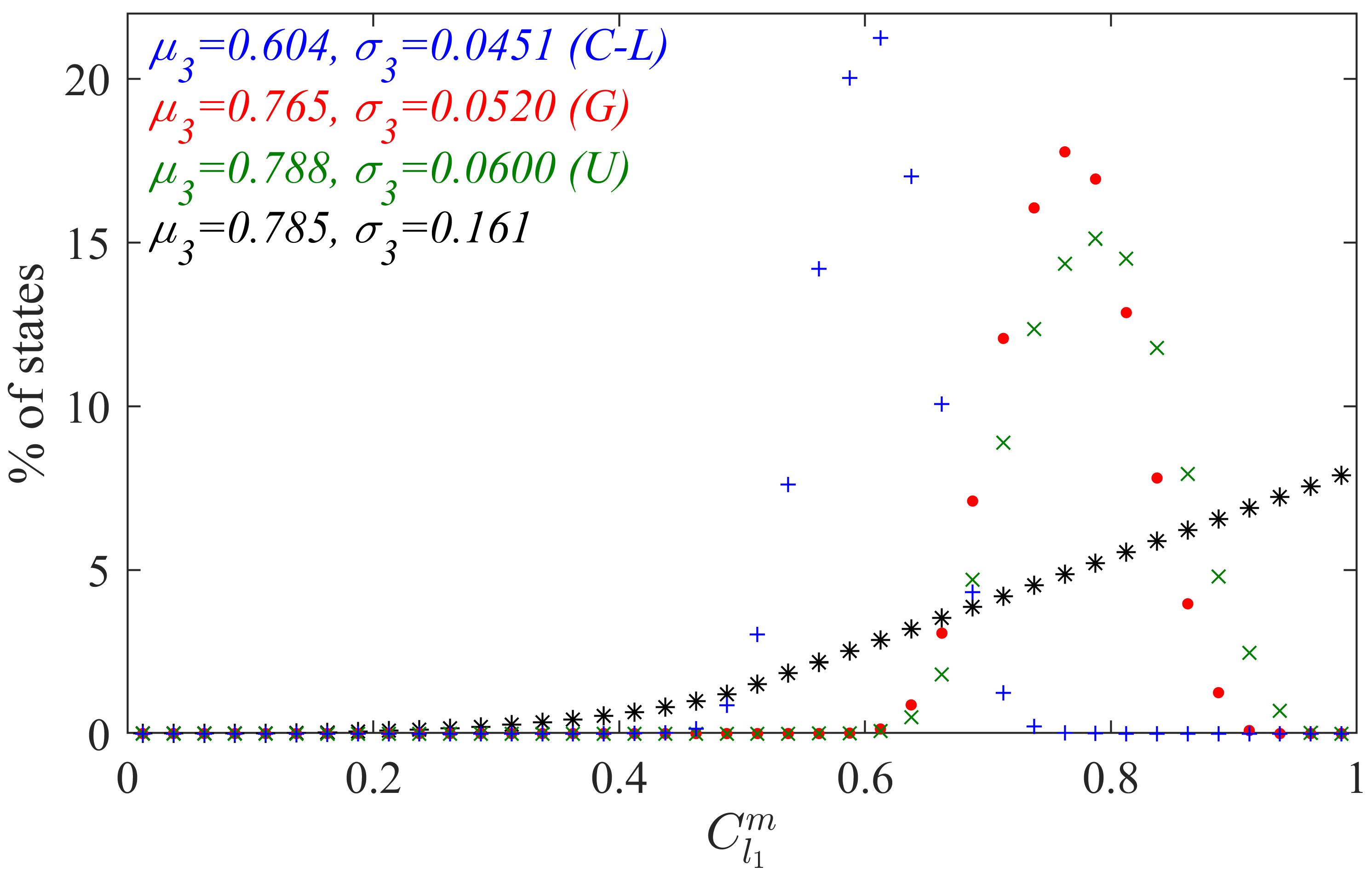}
    \caption{Inhibition of spread of quantum coherence of typical qutrits in response to disorder. The considerations are the same as in Fig.~\ref{l1n2}, except that the states are qutrits.
    The skewnesses for the ordered case and the disordered cases from uniform, Gaussian, and Cauchy-Lorentz distributions are
    $s=-0.865$, 
    $s_U=0.0479$, $s_G=-0.0139$, $s_{C-L}=-0.0437$ respectively.}
    
    
  \label{l1n3}
\end{figure}

\begin{figure}[!ht]
  \centering
    \includegraphics[width=\linewidth]{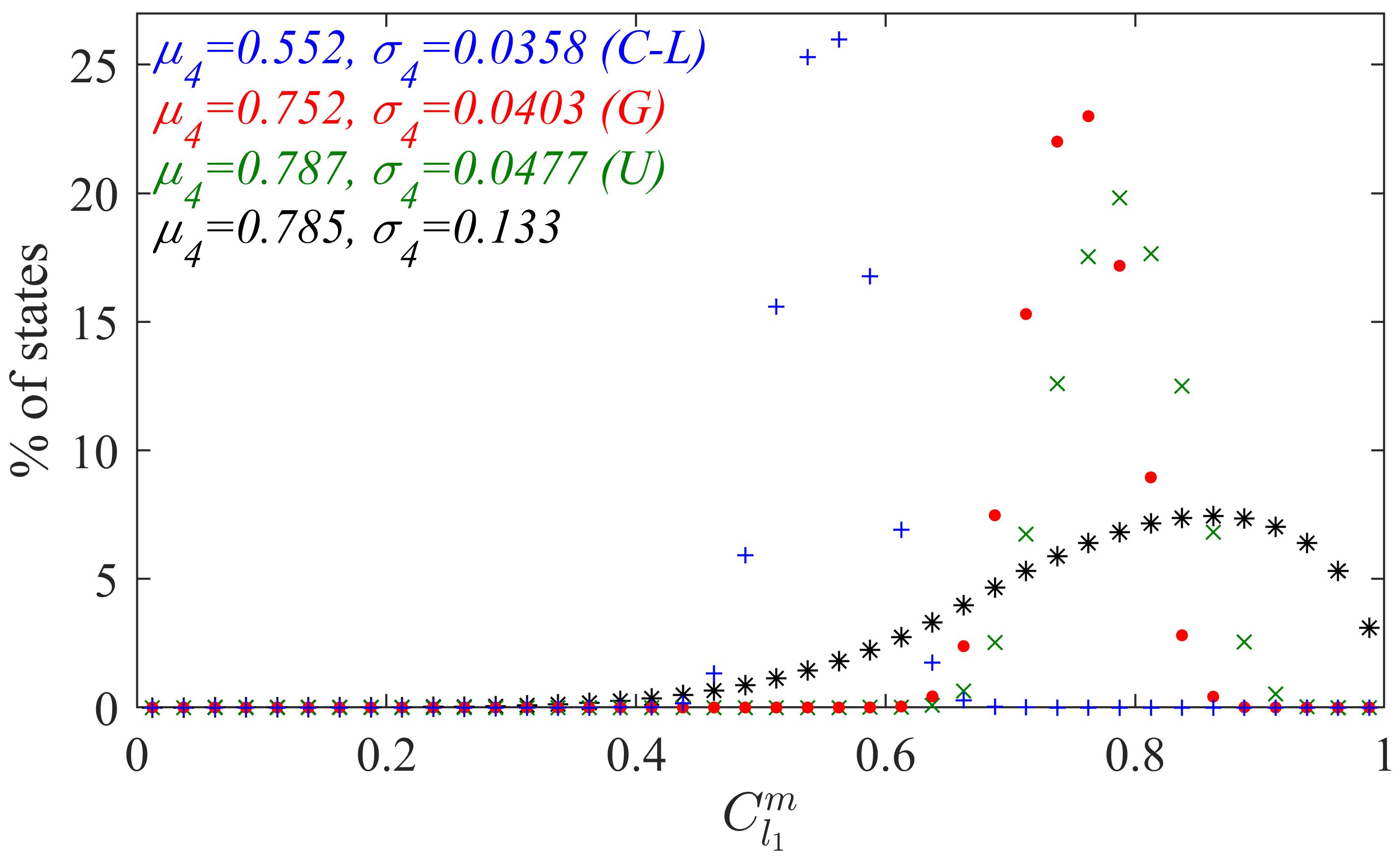}
    \caption{Inhibition of spread of quantum coherence of typical four-dimensional quantum states in response to disorder. The considerations are the same as in Fig.~\ref{l1n2}, except that the quantum states are four-dimensional.
    The skewnesses for the ordered case and the disordered cases from uniform, Gaussian, and Cauchy-Lorentz distributions are
    $s=-0.737$, 
    $s_U=-0.0316$, $s_G=-0.0589$, $s_{C-L}=0.0227$ respectively.}
      \label{l1n4}
\end{figure}

 The effect of disorder on the quantum coherence spread of typical quantum states of higher dimensions is similar to its impact on qubits. 
 Figs.~\ref{l1n3} and~\ref{l1n4} show the effect of disorder on the quantum coherence spread of typical qutrits and typical four-dimensional quantum states.

\begin{figure}[!ht]
  \centering
    \includegraphics[width=\linewidth]{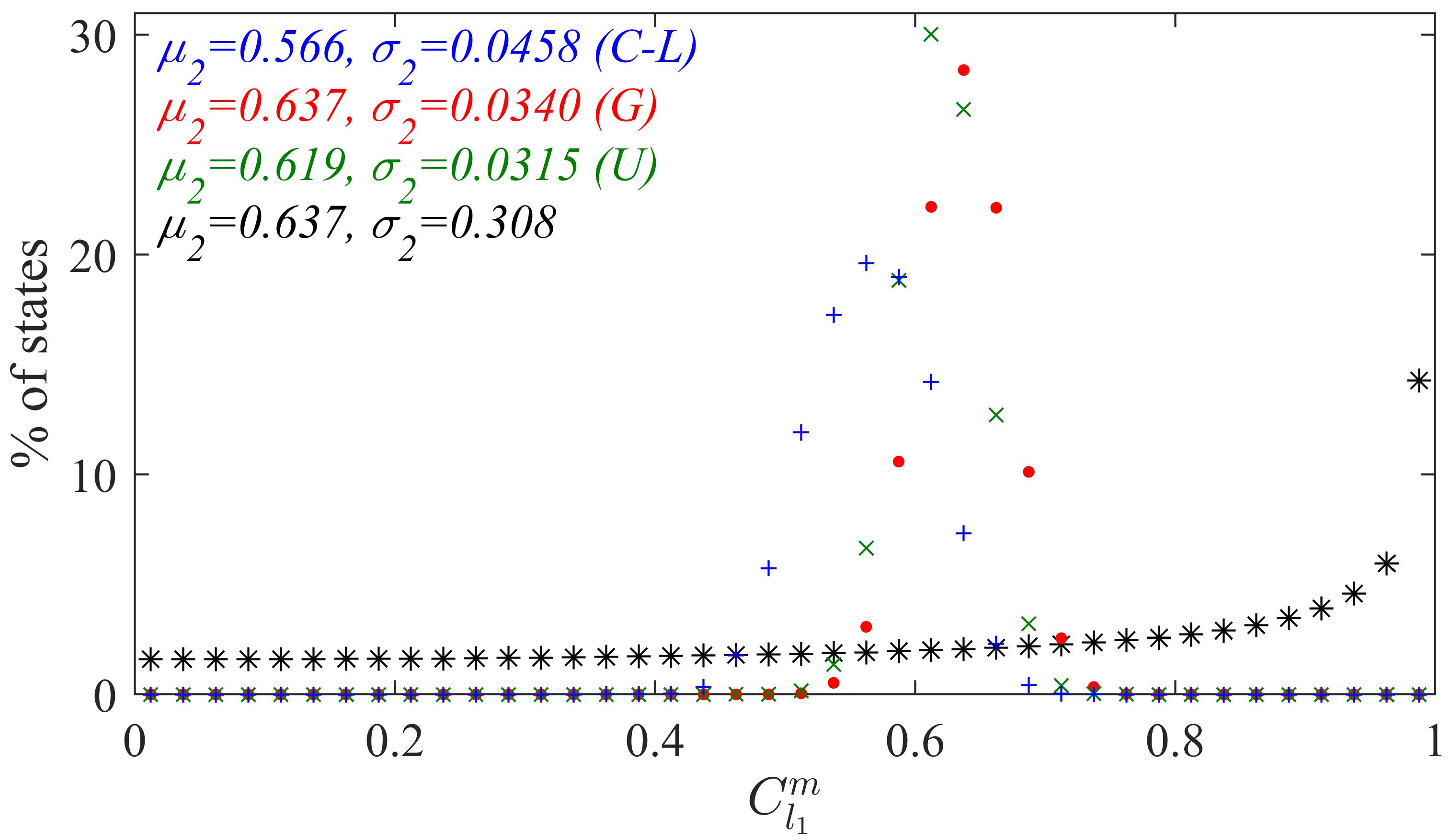}
    \caption{Inhibition of spread of quantum coherence of typical rebits in response to disorder. The considerations are the same as in Fig.~\ref{l1n2} except the fact that we consider rebits in place of qubits.
    The skewnesses for the ordered case and the disordered cases from uniform, Gaussian, and Cauchy-Lorentz distributions are
    $s=-0.498$, 
    $s_U=-0.0564$, $s_G=-0.0558$, $s_{C-L}=-0.0537$ respectively.}
  \label{rebit}
\end{figure}

\begin{figure}[htpb]
  \centering
    \includegraphics[width=\linewidth]{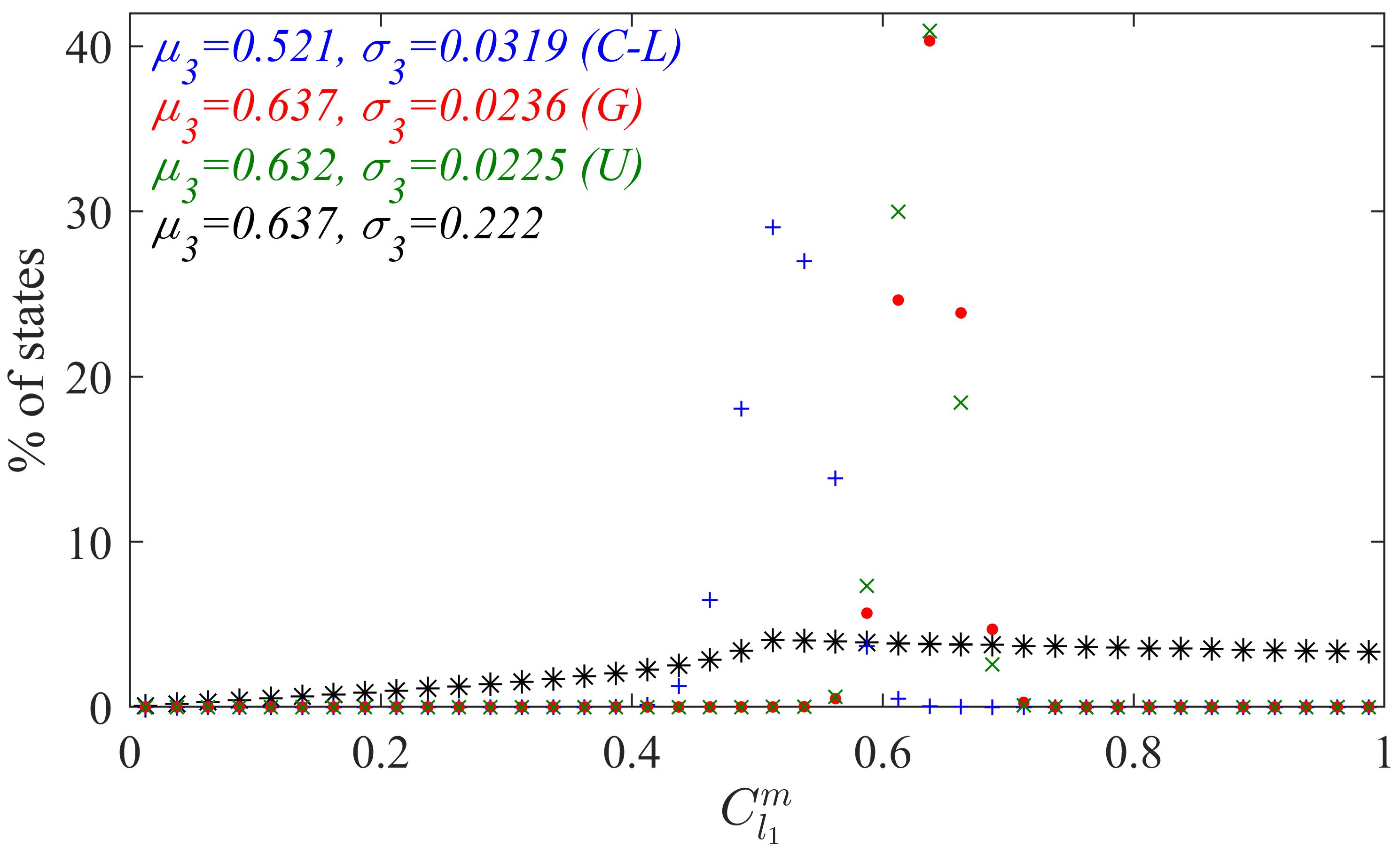}
    \caption{Inhibition of spread of quantum coherence of typical retrits in response to disorder. The considerations are the same as in Fig.~\ref{l1n2} except the fact that we consider retrits in place qubits. 
    The skewnesses for the ordered case and the disordered cases from uniform, Gaussian, and Cauchy-Lorentz distributions are
    $s=-0.375$, 
    $s_U=-0.0292$, $s_G=-0.0666$, $s_{C-L}=-0.0487$ respectively.
    }
  \label{rebit_3_l1n}
\end{figure}

 For completeness, we have also studied the effect of disorder on quantum coherence of typical redits. We again notice that disorder inhibits its spread in any dimension. Figs.~\ref{rebit} and~\ref{rebit_3_l1n} show the inhibition of  spread of typical rebits and retrits for disorders introduced from Gaussian, uniform, and Cauchy-Lorentz distributions with semi-interquartile ranges $\gamma_G=\gamma_U=\gamma_{C-L}=\frac{1}{2}$ respectively. 
 {The qudits and redits are vectors in different vector spaces, and therefore, the behavior of quantum coherences in the two spaces is expected to differ. They are so, but there is significant qualitative similarity. The rates of reduction of the spreads of the distributions of typical modified $l_1$-norms of quantum coherence for qudits and redits have similar behavior but are not the same.}

\section{Conclusion}
\label{qc_sec_5}

We analyze  the average quantum coherence (as quantified by the $l_1$-norm of quantum coherence) of typical redits and qudits, and its response to disorder infusion.
Along with the average quantum coherence of typical states, we have also studied the relative frequency distribution of quantum coherence for vectors of real- and complex-field Hilbert spaces (``redits'' and ``qudits'').
We found that with increasing dimension, the distributions have less spread and become more symmetric. More precisely, the standard deviations and left skewnesses of the relative frequency distributions were found to decrease exponentially with increase in dimension. Moreover, we found that introduction of disorder in the state parameters 
inhibits the spread of the relative frequency distributions. Intuitively,  it may appear that if we disturb the state parameters, the distribution of the
quantum coherence will also be disturbed: the mean will change, and the spread will increase. We found that the exact opposite happens for the spread: it decreases. The more substantial the perturbation, stronger the reduction of the standard deviation. We observed that when perturbed, there is a significant probability for a state with a quantum coherence lower (higher) than the average quantum coherence (of the ordered case)  to jump to one with more (less) quantum coherence than in the parent state. Here, the average quantum coherence is the mean modified $l_1$-norm of Haar uniformly distributed states in the entire relevant Hilbert space, which is 
\(\pi/4\) for qudits and 
\(2/\pi\) for redits. This phenomenon results in a clustering effect which forces the disorder inflicted system to have a low
standard deviation in the quantum coherence distribution. {We claim that} this feature of inhibition of the spread of typical quantum coherence is generic, irrespective of the strength and type of disorder and dimension of the vector space.
{We wish to prove that the results obtained in the manuscript are actual for arbitrary types of disorder distributions. Unfortunately, however, we were unable to verify that statement. Therefore, we used the Gaussian and uniform distributions as examples, as these are widely used in the community to consider disorder. We added the Cauchy-Lorentz distribution to our ``list'', as this is in certain ways very different from Gaussian and uniform distributions. In particular, it does not have a well-defined mean (although it does have a Cauchy mean) and standard deviation. The results support our claim that inhibition in response to disorder is a generic feature for typical pure quantum states.}
Moreover, introduction of disorder results in the quantum coherence distribution to become more symmetric. 

Disorder is ubiquitous in quantum systems, and quantum coherence is a valuable resource in understanding quantum systems and quantum devices. We repeated the exercise for real vector spaces and found qualitatively similar results for completeness. We believe the results will be significant in a broad spectrum of areas, including quantum systems and devices.

\acknowledgements
 US acknowledges partial support from the Department of Science and Technology, Government of India through the QuEST grant (grant number DST/ICPS/QUST/Theme-3/2019/120). 

\bibliography{5th_paper}
\end{document}